\input{aipcheck}

\documentclass[
   ,final            
  ]
  {aipproc}

\layoutstyle{8x11double}

\newcommand\fverb{\setbox\fverbbox=\hbox\bgroup\verb}
\newcommand\fverbdo{\egroup\medskip\noindent%
			\fbox{\unhbox\fverbbox}\ }
\newcommand\fverbit{\egroup\item[\fbox{\unhbox\fverbbox}]}
\newbox\fverbbox

\newcommand{\mpl}{M_{\rm Pl}}

\newcommand{\del}{\partial}
\newcommand{\phie}{\phih_{\rm{end}}}
\newcommand{\phih}{\phi}
\newcommand{\gh}{g}
\newcommand{\lambdah}{\lambda}
\newcommand{\Mh}{M}

\begin{document}

\title{Inflation in the nonminimal theory with `$K(\phi)R$' term}


\author{Seong Chan Park}{
  address={
FRDP, Department of Physics and Astronomy,
Seoul National University,
Seoul, Korea}}

\begin{abstract}
  A class of inflationary models with the nonminimal coupling term `$K(\phi)R$' is considered.
  We show that the successful inflation can take place at large field value limit once the ratio between the square of the nonminimal coupling term and the  potential for the scalar goes asymptotically constant ($V(\phi)/K^2(\phi) \rightarrow Const$) \footnote{Talk given at 16th International Conference on Supersymmetry and the Unification of Fundamental Interactions (SUSY08), Seoul, Korea, 16-21 Jun 2008. This talk is based on the paper ~\cite{Park:2008hz}}.
\end{abstract}

\maketitle


It is widely accepted that the idea of inflation~\cite{inf}
is the best solution to many cosmological problems such as flatness, homogeneity and isotropy of
the observed universe~\cite{books}.
In models of particle physics models of inflation, it took place essentially due to a scalar field,
the inflaton field, whose potential is so flat that the inflaton can roll down only very slowly~\cite{Lyth:1998xn}.
Under such a `slow-roll' condition, the curvature perturbation is produced nearly scale invariant way
and this feature is precisely confirmed by the measurements of the anisotropies of the
CMB and the observations of the large scale structure~\cite{obs}.
The biggest question is the origin of the inflaton field itself and the form of its nearly flat potential.

Very Recently Bezrukov and Shaposhnikov (BS) reported an intriguing possibility that
the standard model with an additional non-minimal coupling term of the Higgs field ($H$) and
the Ricci scalar ($\sim a |H|^2 R$) can give rise to inflation ~\cite{Bezrukov:2007ep}
without introducing any additional scalar particle in the theory. \footnote{There were models of chaotic
inflation with nonzero $a$ suggested in literatures in various different contexts
~\cite{Spokoiny:1984bd,Salopek:1988qh,Kaiser:1994vs,Komatsu:1999mt,Futamase:1987ua,Fakir:1990eg,Libanov:1998wg}}.
The new thing that the BS showed was that the ``physical Higgs potential'' in Einstein frame is indeed nearly flat at the large
field value limit and fit the COBE data $U/\varepsilon=(0.027 \mpl)^4$ once the ratio
between the quartic coupling of the Higgs field ($\lambda$) and the non-minimal coupling constant ($a$)
 is chosen to be small as $\sqrt{\lambda/a^2}\sim 10^{-5}$.

Here we found several interesting questions in this model. What is the underlying reason why the theory can work. What is the role of the nonminimal coupling term? What is the condition for the nonminimal term to fit the real data of cosmological observations?
To address this question, we would generalize the case of BS by taking more generic form of the nonminimal coupling and look for the required condition for the asymptotically flat potential. It is certainly worthwhile to consider the generalization since we could understand the underlying structure of the theory more closely \cite{Park:2008hz}.


Let us start from the model with non-minimal coupling $K(\phih)$ and the scalar potential $V(\phih)$.
The action in Jordan frame is given as
\begin{eqnarray}
 S=\int d^4 x \sqrt{-\gh}(-\frac{\Mh^2+K(\phih)}{2}{R}+
 \frac{1}{2}(\del \phih)^2-V(\phih)).
\end{eqnarray}
One should notice that if we take  $K(\phih)= a |\phih|$ and $V(\phih)=\lambda (|\phih|^2-v^2)^2$, the action is reduced to
the original action which is taken by BS. Here we are considering a generalized version of the potential.
The Einstein metric is obtained as
\begin{eqnarray}
 \gh_{\mu\nu}=e^{-2\omega}g^{E}_{\mu\nu},\qquad
 e^{2\omega}:=\frac{\Mh^2+K(\phih)}{\mpl^2}.
\end{eqnarray}
By the conformal transformation, we get the action in the Einstein frame as follows.
\begin{eqnarray}
 \int d^4x\sqrt{-g_{E}}(
-\frac{\mpl^2}{2}R_{E}
+\frac34\frac{e^{-4\omega}}{\mpl^2}K'(\phih)^2(\del \phih)^2 \nonumber \\
+\frac12 e^{-2\omega}(\del \phih)^2
-e^{-4\omega}V(\phih)).
\end{eqnarray}
It is convenient to redefine the scalar field and normalize the kinetic term canonically.
\begin{eqnarray}
 \frac{dh}{d\phih}=\sqrt{
\frac{\mpl^2}{\Mh^2+K(\phih)}
+\frac32\frac{\mpl^2}{(\Mh^2+K(\phih))^2}K'(\phih)^2
}.\label{dhdphi}
\end{eqnarray}
Now the physical scalar potential in the Einstein frame is written as
\begin{eqnarray}
 U=\frac{\mpl^4}{(\Mh^2+K(\phih))^2}V(\phih). \label{U}
\end{eqnarray}
Here we could read out the general condition for the flat potential
at the large field value:
\begin{eqnarray}
\lim_{\phih\rightarrow \infty}\frac{V}{K^2} = Const >0.
\label{condition}
\end{eqnarray}
since $U \sim \frac{V}{K^2}$. The condition $K(\phih) \gg \Mh^2$ for $\phih \gg \Mh$
is required for the potential to be bounded from below and the location of the global minimum
is well localized around the small field value.
Even though the condition in eq. \ref{condition} actually determines the flatness of the potential at the large field value,
it is not necessarily required in generic inflation models. Depending on the shape of the potential, it might still be
possible to have sufficient time of exponential expansion for some {\it finite} region of field value $\phi$.
The result is certainly applicable for monotonic potentials, for example, monomial potentials which will be considered
below in great detail.


Now let us consider the case when $K(\phih)$ is a monomial as
\begin{eqnarray}
 K(\phih)=a \phih^m,
\end{eqnarray}
where $a$ is a dimensionful constant in general.
In order to get the flat potential in large $\phih$ region in Einstein frame, the original scalar potential in Jordan frame should be written as
\begin{eqnarray}
 V=\frac{\lambdah}{2m}\phih^{2m}.
\end{eqnarray}
In this case, $U$ is written as
\begin{eqnarray}
 U=\frac{\mpl^4\lambdah}{2m a^2}\left(1+\frac{\Mh^2}{a}\phih^{-m}\right)^{-2}
 \label{potentialm}
\end{eqnarray}

The slow roll parameters are defined by using the scalar potential in Einstein frame \ref{U} and the canonically normalized scalar field $h$ as
\begin{eqnarray}
 \varepsilon=\frac{\mpl^2}{2}\left(\frac{\del U/\del h}{U}\right)^2,
 \qquad
 \eta=\mpl^2\frac{\del^2 U/\del h^2}{U}.
\end{eqnarray}

In our model these parameters are calculated in large $\phih$ region, using eqs.
\ref{potentialm}, as
\begin{eqnarray}
 \varepsilon=
 \left\{
   \begin{array}{ll}
     \frac{2M}{a}\left(\frac{M}{\phih}\right)^3, & \hbox{$m=1$;} \\
     \frac{4}{3a^2(1+1/(6a))}\left(\frac{M}{\phih}\right)^4, & \hbox{$m=2$;} \\
     \frac{4M^{-2m+4}}{3a^2}\left(\frac{M}{\phih}\right)^{2m}, & \hbox{$m\ge 3$.}
   \end{array}
 \right.
,\\
 \eta=
 \left\{
   \begin{array}{ll}
     -3\left(\frac{\Mh}{\phih}\right)^2, & \hbox{$m=1$;} \\
     -\frac{4}{3a(1+1/(6a))}\left(\frac{\Mh}{\phih}\right)^2, & \hbox{$m=2$;} \\
     -\frac{4\Mh^{2-m}}{3a}\left(\frac{\Mh}{\phih}\right)^m, & \hbox{$m\ge 3$.}
   \end{array}
 \right.
\label{epsilon-eta}
\end{eqnarray}

The end of inflation is $\varepsilon=1$. The values of $h$ and $\phih$ at this point are denoted by $h_{\rm end}$ and $\phie$ respectively. In the slow roll inflation the number of e-foldings is expressed as
\begin{eqnarray}
 N=\frac{1}{\mpl^2}\int^{h_0}_{h_{\rm{end}}}\frac{U}{\partial U/\partial h}.
\end{eqnarray}
In our model $N$ is calculated as
\begin{eqnarray}
 N=
\left\{
\begin{array}{ll}
 \frac{1}{4M^2}(\phih_0^2-\phie^2), & \hbox{$(m=1)$}\\
 \frac{3}{4} a\left(1+\frac{1}{6a}\right)\frac{1}{M^2}(\phih_0^2-\phie^2), & \hbox{$(m=2)$}\\
 \frac{3}{4} a\frac{1}{\Mh^{2}}(\phih_0^m-\phie^m) , & \hbox{$(m \ge 3)$}
\end{array}
\right.
\end{eqnarray}
In order to get $60$ e-foldings, we should solve $N=60$ and get $\phih_{60}$.
Let us assume $\phih_{60} \gg \phie^2$. Then we obtain the value $\phih_{60}$ as
\begin{eqnarray}
  \phih_{60}=
\left\{
  \begin{array}{ll}
    2\sqrt{N}M, & \hbox{($m=1$)} \\
    \frac{2 \sqrt{N}M}{\sqrt{3a(1+1/(6a))}}, & \hbox{($m=2$)} \\
    \left(\frac{4 N}{3a}M^2\right)^{1/m}, & \hbox{($m\ge 3$).}
  \end{array}
\right.
\label{phi60}
\end{eqnarray}

The spectral index $n_s$ and the tensor-to-scalar ratio $r$ can be calculated as
\begin{eqnarray}
 n_s=1-6\varepsilon+2\eta|_{\phih=\phih_{60}},\qquad
 r=16\varepsilon|_{\phih=\phih_{60}}.
\end{eqnarray}
In our model, these values are expressed (using eq.\ref{epsilon-eta} and eq.\ref{phi60}) as
\begin{eqnarray}
 n_s=
\left\{
\begin{array}{ll}
 1-\frac{3}{2a_0N^{3/2}}-\frac{3}{2N}, & \hbox{$(m=1)$}\\
 1-\frac{9(1+1/(6a_0))}{2N^2}-\frac{2}{N}, & \hbox{$(m=2)$}\\
 1-\frac{9}{2N^2}-\frac{2}{N}, & \hbox{$(m\ge 3)$}
\end{array}\right.
,\qquad \\
r=
\left\{
\begin{array}{ll}
 \frac{4}{a_0N^{3/2}}, & \hbox{$(m=1)$}\\
 \frac{12(1+1/(6a_0))}{N^2}, & \hbox{$(m=2)$}\\
 \frac{12}{N^2} , & \hbox{$(m\ge 3)$}
\end{array}\right.
\end{eqnarray}
where the dimensionless parameter $a_0$ is defined as
\begin{eqnarray}
 a_0=a M^{m-2}.
\end{eqnarray}

\begin{figure}
  \includegraphics[width=.35\textheight]{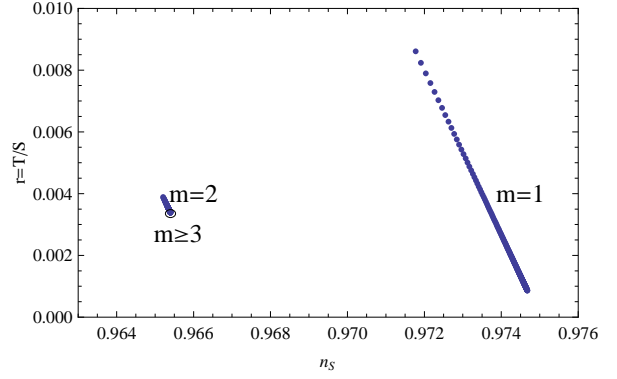}
  \caption{The spectral index $n_s$ and the tensor-to-scalar perturbation ratio $r$
are depicted in one plot for various values of $a_0$ and the power of the
non-minimal coupling $m$ in $K(\phi)\sim \phi^m $. }
\label{rnplot}
\end{figure}

In fig.\ref{rnplot} we plotted the spectral index ($n_S$) and the tensor-to-scalar
perturbation ratio ($r$) for varying $a_0$ and fixed $N=60$. For $m=1$ and $m=2$,
the spectral index becomes larger but the tensor-to-scalar ratio becomes smaller.
For {\it large} $a_0\simeq 4\pi$,
the values of the spectral index and the tensor-to-scalar ratio are saturated to
$0.9745 (0.965)$ and $0.0007(0.003)$ for $m=1(m\geq 2)$, respectively. Notice that
when $m\geq 3$, the spectral index and $r$ are independent of $a_0$ and
given as $0.965$ and $0.003$, respectively. It is depicted by a circle at the tip
of the plot for $m=2$.

Another observable is the amplitude of the scalar perturbation.
\begin{eqnarray}
 \delta_{H}=\frac{\delta \rho}{\rho}\cong \frac{1}{5\sqrt{3}H}\frac{U^{3/2}}{\mpl U'}=1.91\times 10^{-5}.
\end{eqnarray}
This gives a constraint for the parameters
\begin{eqnarray}
 \frac{U}{\epsilon}=(0.027\mpl)^4.
\end{eqnarray}
In our model, the constraint is written, with dimensionless parameter
 $\lambda_0=\lambda \Mh^{2m-4}$, as follows.
\begin{eqnarray}
 \begin{array}{ll}
  \sqrt{\frac{\lambda_0}{a_0}}&\simeq 2.3\times 10^{-5},~~ (m=1)\\
  \sqrt{\frac{\lambda_0}{a_0^2(1+1/(6a_0))}}&\simeq 2.1\times 10^{-5},~~ (m=2)\\
  \sqrt{\frac{\lambda_0}{a_0^2}}&\simeq 1.5\times 10^{-5}\sqrt{m},~~ (m\ge 3).\\
 \end{array}
\label{condition2}
\end{eqnarray}
One should note that $\sqrt{\frac{\lambda_0}{a_0^2}}\sim 10^{-5}$ is
universally required to fit the observational data for general values of $m$.
However this is weird since the quartic coupling has to be extremely small
$\lambda \sim 10^{-10} a_0^2$ as we already noticed in the case with $m=2$.

Now let us summarize the paper. We study the inflationary scenarios based on the theory with non-minimal coupling of a
scalar field with the Ricci scalar ($\sim K(\phi)R$). Taking conformal transformation, the resultant
scalar potential in the Einstein frame is shown to be flat at the large field limit if the condition
in eq.\ref{condition} is satisfied. This is one of the main result of this paper.
This class of models gets constraints from the recent cosmological observations
of the spectral index, tensor-to-scalar perturbation ratio as well as the amplitude of the potential.
We explicitly considered the monomial cases $K \sim \phi^m$ and found that this class of models are indeed good
agreement with the recent observational data: $n_S \simeq 0.964-0.975$ and $r \simeq 0.0007- 0.008$ for any value of $m$.
In fig.\ref{rnplot}, the predicted values for $n_S$ and $r$ are depicted. We explicitly read out the condition for
fitting the observed anisotropy of the CMBR by which essentially the amplitude of the potential is determined. The condition
does not look natural ($\sqrt{\lambda/a^2}\sim 10^{-5}$) at the first sight but we may understand this seemingly unnatural value
once we embed the theory in higher dimensional space-time. Details of higher dimensional embedding of the theory and
possible solution to the smallness of $\sqrt{\lambda/a^2}$ will be given in separate publication \cite{large volume}.

\begin{theacknowledgments}
  This talk is based on a work done in collaboration with Satoshi Yamaguchi of SNU.
\end{theacknowledgments}

\end{document}